\documentclass[%
reprint,
footinbib,
aps,
superscriptaddress,
]{revtex4-1}

\usepackage{hyperref}
\hypersetup{colorlinks=true, linkcolor=blue, urlcolor=blue, citecolor=blue}
\usepackage{graphicx}
\usepackage{amsmath} \usepackage{amssymb}
\usepackage{accents} \usepackage{mathrsfs} \usepackage{setspace}
\usepackage{color}
\definecolor{dgreen}{RGB}{00, 120, 00} \definecolor{dblue}{RGB}{00, 00, 220}
\definecolor{lgreen}{RGB}{46, 139, 87} 
\usepackage{ulem}

\newcommand{\bs}[1]{\boldsymbol{#1}} \newcommand{\ul}[1]{\underline{#1}}
\newcommand{\tr}[1]{\mathrm{Tr}\left[#1\right]}  
\newcommand{\re}[1]{\mathrm{Re}\left[#1\right]}

\newcommand{\ut}[1]{\undertilde{#1}} 
\newcommand{\ub}[1]{\underline{#1}} 
\newcommand{\ve}[0]{\varepsilon} \newcommand{\vphi}[0]{\varphi}

\newcommand{\dw}[0]{($d+id'$)-wave }
\newcommand{\pw}[0]{($p+ip'$)-wave }

\begin{document} 
\title{Chiral Current Inversion Induced by Flat-Band Andreev Bound States}

\author{Shu-Ichiro Suzuki}
\affiliation{Faculty of Science and Technology and MESA+ Institute for Nanotechnology, University of Twente, 
7500 AE Enschede, The Netherlands}

\author{Alexander A. Golubov}
\affiliation{Faculty of Science and Technology and MESA+ Institute for Nanotechnology, University of Twente, 
7500 AE Enschede, The Netherlands}

\author{Matthias Eschrig}
\affiliation{University of Greifswald, Institute of Physics, 17489 Greifswald, Germany}

\date{\today}

\begin{abstract}
  We study the spontaneous chiral surface current circulating in a three-dimensional disk of a chiral superconductor (SC) utilizing quasiclassical Eilenberger theory. We obtain spatial profiles of the chiral current for both a ($d_{zx} + i d_{yz}$)-wave and a ($p_{x} + i	p_{y}$)-wave SCs (where the top and bottom surfaces of the disk are perpendicular to the $z$-axis).  Whereas the chiral current for a ($p_{x} + i p_{y}$)-wave SC does not depend on $z$, a reversal of the chiral current takes place at the top and bottom surfaces in the case of a ($d_{zx} + i d_{yz}$)-wave SC.  In this latter case, flat-band Andreev bound states appear at the top and bottom surfaces in addition to the chiral surface states at the lateral surface. The chiral current reversal is explained in terms of a hybridization between the two types of Andreev bound states. As a result, the magnetic field around the disk differs drastically between the two cases.
\end{abstract}

\pacs{pacs}

\thispagestyle{empty}

\maketitle

\textit{Introduction.}--Chiral surface current (CSC), the spontaneous
surface supercurrent appearing due to spontaneous chiral symmetry
breaking, is one of the most distinct features of chiral
superconductors (SCs) \cite{Matsumoto_JPSJ_1999, Furusaki_PRB_2001,
Stone_04, Nagato_11, Sauls_PRB_2011, Bakurskiy_14, Lederer_PRB_14,
suzuki_16} and chiral superfluids (SFs) \cite{Leggett_RMP_75,
Mermin_Physica_77, Hall_PRL_85, Kita_JPSJ_1998, Tsutsumi_PRB_12,
Tsutsumi_JPSJ_12, Mizushima_JPSJ_16, Tada_PRB_18}.  The non-zero
orbital angular momenta of the Cooper pairs give rise to a circulating
supercurrent along the lateral surfaces of a sample. Although the
presence of the CSC (or the associated spontaneous magnetization)
could be considered a direct proof of the chiral superconductivity, it
has never been observed in any material that is a candidate for chiral
SC \cite{Moler_05, Nelson_07}.
Even in the simplest chiral SF, the $^3$He A-phase, the intrinsic
angular momentum, which is generated by the chiral surface mass
current, has been a controversial issue since its discovery. 
Chiral \textit{superconductivity} is even more intricate.  The
spontaneous time reversal symmetry (TRS) breaking associated with the
superconducting transition has been observed by, for example,
muon-spin rotation/relaxation ($\mu$SR) measurements.  However, even
in those TRS-breaking superconducting states, the CSC has never been
experimentally observed. 
Missing observation of the CSC stands as a serious open problem.  
A comprehensive understanding of the nature and origin of chiral
supercurrents in terms of the underlying spectral properties is
fundamentally essential for unraveling the complexities of
unconventional SCs.

Interlayer Cooper pairing [i.e., ($d_{zx}+id_{yz}$)-wave pairing]
could constitute a mechanism for layered chiral SCs such as
URu$_2$Si$_2$ \cite{Kasahara_PRL_07, Kasahara_NJP_09,
Kittaka_JPSJ_16}, Sr$_2$RuO$_2$ \cite{maeno_94, maeno_03,
Pustogow_Nature_19, Agterberg_PRR_2020, Suzuki_PRB_20, Grinenko_20,
Grinenko_21, Ikegaya_PRR_21, Suzuki_PRR_22, Yuri_PRR_2022}, and
LaPt$_3$P \cite{LaPt3P_NatCommun_21}.  Such an even-parity chiral
pairing explains the spontaneous TRS-breaking and the spin-singlet
behaviour, simultaneously. 
The ($d_{zx}+id_{yz}$)-wave SC hosts chiral surface states responsible
for the CSC appearing at its lateral surface. Furthermore, it has
flat-band Andreev Bound states (ABSs) at its top and bottom surfaces
as a consequence of interlayer pairing \cite{Suzuki_PRR_22}. The
physics of these two types of surface states have separately been
discussed in individual contexts.  Recent investigations have
demonstrated a substantial influence of the flat-band ABSs on the
supercurrent 
\cite{Higashitani_JPSJ_97, 
Asano_PRL_11, 
Suzuki_PRB_14,
Suzuki_PRB_15, 
Bernardo_PRX_15, 
Fogelstrom_NP_15,
Linder_PRL_16, 
Lofwander_NC_18,
Krieger_PRL_20,
Yoshida_PRR_22}. In particular, at the edge between the lateral and
top surfaces, the CSC is expected to be modified by the interaction
between the chiral surface states and the flat-band ABSs.  In this
Letter, we demonstrate an intriguing phenomenon: the inversion of the
CSC at the edges, a result we attribute to the interaction between
flat-band ABSs localized at the top/bottom surfaces and chiral Andreev
bound states at the lateral surfaces. This inversion, notably induced
by these ABSs, leads to a significant reduction in both the total CSC
and the spontaneous magnetization within finite-size samples.

\begin{figure}[b]
	\includegraphics[width=0.48\textwidth]{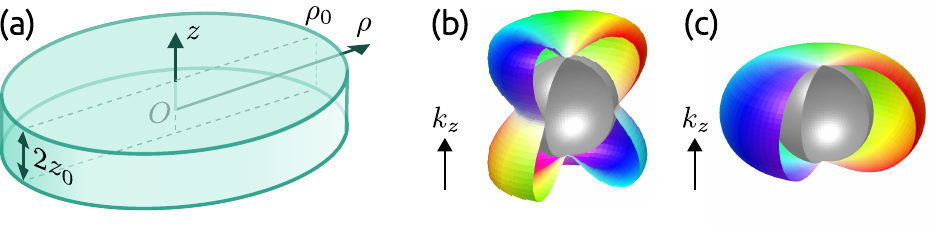}
	\caption{(a) Superconducting disk. 	
  (a,b) Gap structures of ($d+id'$)- and ($p+ip'$)-wave SCs. 
  The color coding represents the phase of $\Delta(\bs{k})$ in the
homogeneous limit.}
  \label{fig:Sche}
\end{figure}

Using the quasiclassical theory of superconductivity, we study the CSC
in a mesoscopic three-dimensional disk made of ($d_{zx}+id_{yz}$)-wave
[$d+id'$-wave] or ($p_x+ip_y$)-wave [$p+ip'$-wave] superconductor 
(See Fig.~\ref{fig:Sche}). 
We demonstrate that, in the ($d+id'$)-wave case, the chiral
lateral-surface currents at the top and bottom edges flow in  opposite
direction as compared with the CSC far from the edges.  Upon
comparison between the results observed in ($d+id'$)- and
($p+ip'$)-wave SCs, we will show that the flat-band ABSs are the
origin of the CSC inversion. 
Given the current density, we calculate the spatial distribution of
the spontaneous magnetic field. The spontaneous field of the
($d+id'$)-wave SC is much smaller than that of the ($p+ip'$)-wave SC.
Although the counter chiral currents (CCC) are localized at the
top/bottom edges, the contribution from the local inverted currents
drastically changes the local field near the top surface, where a
scanning SQUID is applied to measure the spontaneous magnetic field. 

\textit{Quasiclassical Eilenberger theory.}--We examine the effects of
the flat-band ABSs utilizing the quasiclassical Eilenberger theory
\cite{Eilenberger} in a three-dimensional disk geometry
[Fig.~\ref{fig:Sche}(a)].  The radius and thickness of the disk are
denoted by $\rho_0$ and $2z_0$, respectively.  The pair potential has
either ($d+id'$)- or ($p+ip'$)-wave symmetry
[Fig.~\ref{fig:Sche}(b,c)]. 
The quasiclassical Green's functions (QGFs) obey the Eilenberger
equation which is valid in the weak-coupling limit, 
\begin{align}
  & i \bs{v}_F \cdot \bs{\nabla} \tilde{g}
	+ \left[ \, i\omega_n \tilde{\tau}_3+\tilde{\Delta},~\tilde{g} \right]_-
	= 0, 
	\label{eq:Eilen}
	\\[2mm]
	& \tilde {g} 
	= \left( \begin{array}{rr}
	     g  &  
	     f  \\[1mm]
	-\ub{f} & 
	-    g  \\
	\end{array} \right), 
	\hspace{4mm}
	 \tilde {\Delta} = \left( \begin{array}{cc}
	 0&  
	 i \Delta \\[1mm]
	 i \Delta^* & 
	 0\\
	\end{array} \right), 
\end{align}
where 
$\tilde{g}(\bs{r},\bs{k},i\omega_n)$ is the QGF, 
$\Delta(\bs{r},\bs{k})$ is the pair potential, and 
$\bs{k}=(
\sin \theta_k \cos \phi_k, 
\sin \theta_k \sin \phi_k, 
\cos \theta_k
)$ indicates the direction
of the Fermi momentum. 
In this Letter, $(\tilde{\cdot})$ means a matrix in
Nambu space: The identity and Pauli matrices are 
denoted 
by $\tilde{\tau}_0$ and $\tilde{\tau}_\nu$,
respectively,
with $\nu \in \{1, 2, 3 \}$. 
Underlined functions are defined by 
$K (\bs{r},\bs{k},i \omega_n) 
= s_\nu \ub{K}^* (\bs{r},-\bs{k},i \omega_n)$ with $s_\nu = -1$
($+1$) for the singlet (triplet) SC. 
The type-II limit is considered. 
The QGF is supplemented by the
normalization condition $
\tilde {g} 
\tilde {g} 
=
\tilde{\tau}_0$. 
Throughout this Letter, we use $\hbar=c=k_B=1$ and
$\bs{r}=(
\rho \cos \vphi, 
\rho \sin \vphi, 
z)$. 

The Eilenberger equation \eqref{eq:Eilen} can be written in terms of
the Riccati parameterization \cite{Schopohl_PRB_95, 
Eschrig_PRB_00, Eschrig_PRB_09}, in which the QGF is given
in terms of the coherence function $\gamma(\bs{r}, \bs{k}, i \omega_n)$, 
\begin{align}
	& \tilde {g} = 
	\frac{2 }{1-\gamma \ul{\gamma}}
	\left( \begin{array}{cc}
	 1 &  
	     \gamma \\[1mm]
	-\ul{\gamma} & 
	-1 \\
	\end{array} \right) - \tilde{\tau}_3. 
\label{eq:Ric-Para}
\end{align}
The equation for $\gamma$ is given by 
\begin{align}
  & \bs{v}_F \cdot \bs{\nabla}  \gamma
	+ 2 \omega_n \gamma
	  - \Delta
		+ \Delta^*
	\gamma^2= 0. 
	\label{eq:Riccati07}
\end{align}
For the homogeneous case $\Delta=\bar\Delta=$const., $\gamma$ is given by its bulk value
$
\bar{\gamma}=
	\bar{\Delta}/ 
	({\omega_n + \sqrt{\omega_n^2 + |\bar{\Delta}|^2} }) 
$.

We consider $p+ip'$ and $d+id'$ order parameters of the form $\Delta = \Delta_\pm
(k_x\pm ik_y)$ and $\Delta =
2\Delta_\pm (k_x\pm ik_y)k_z$. Assuming a dominant
$\Delta_{+}$
component, 
a subdominant 
$\Delta_{-}$
component with 
$e^{2i\varphi}$ 
appears in inhomogeneous systems for
the energetically most stable configuration \cite{SaulsEschrig09}. The
two-component pair potential is thus given by
\begin{align*}
  \Delta = \left\{ 
	\begin{array}{ll}
	(\Delta_+ e^{i\phi_k} + \Delta_- e^{-i\phi_k}e^{2i\vphi}) \sin(2\theta_k)  & \text{for $(d+id')$},\\
  (\Delta_+ e^{i\phi_k} + \Delta_- e^{-i\phi_k}e^{2i\vphi}) \sin \theta_k    & \text{for $(p+ip')$}.
	\end{array}
	\right.
\label{}
\end{align*}
The spatial dependence of $\Delta_\pm(\rho,z)$ is determined by the
self-consistency (gap) equation, 
\begin{align}
	& \Delta_\pm(\bs{r})
	=
	2 \lambda \nu_0 \frac{\pi}{i \beta} \sum_{\omega_n}^{\omega_c}
	\langle V_\pm(\bs{k}') f(\bs{r},\bs{k}',i \omega_n) \rangle, 
\end{align}
with BCS coupling constant
\begin{align}
	& \lambda 
	= \frac{1}{2 \nu_0}
	\left[
		\ln\frac{T}{T_{c0}} 
	+ \sum_{n=0}^{n_c}
	\frac{1}{n+1/2}
\right]^{-1}, 
\end{align}
where 
$ \langle \cdots \rangle
  = \int_0^\pi \int_{-\pi}^\pi \cdots 
	{\sin \theta_k d\phi_k d\theta_k}/{4 \pi}$,
	$\beta=1/T$, $T_{c0}$ is the critical temperature in the homogeneous
	case,
$\nu_0$ is the density of the states (DOS) in the normal state at the
Fermi energy, and $n_c$ is the cutoff integer corresponding to $\omega_c$. 
Throughout this letter, we fix 
$\omega_c = 6 \pi T_{c0}$. 

The attractive potentials $V_\pm$ are given by, 
\begin{align*}
  V_\pm = \left\{ 
	\begin{array}{cc}
		15 e^{\mp \phi_k} \sin (2\theta_k) /8  & \text{for $(d+id')$} \\
	   3 e^{\mp \phi_k} \sin   \theta_k  /2  & \text{for $(p+ip')$} \\
	\end{array}
	\right.
\label{}
\end{align*}

The spontaneous chiral current is calculated from the QGF, 
\begin{align}
	\bs{j}(\bs{r})
	&= ev_F \frac{2 \pi \nu_0}{i \beta} \sum_{\omega_n}^{\omega_c} 
	\langle \bs{k} g(\bs{r},\bs{k},i \omega_n) \rangle, 
  \label{eq:cur_exp1}
	\\
	& = \frac{|e| v_F }{2}
	\int_{-\infty}^{\infty}
	\left \langle \bs{k} \nu_{\bs{k}} 
	\right \rangle \tanh \left( \frac{E}{2T} \right) 
	dE,
  \label{eq:cur_exp2}
	%
\end{align}
where $\nu_{\bs{k}}(\bs{r},E) = 2 \nu_0 \re{g(\bs{r}, \bs{k}, i\omega_n)|_{i \omega_n \to
	E+i\delta}}$ is the angle-resolved density of states (ARDOS), 
$\delta$ is an infinitesimal positive quantity, 
and $e <0$ is the electron charge.

The local field generated by $\bs{j}(\bs{r})$ can be calculated
by the Biot-Savart law, 
\begin{align}
	\bs{h}(\bs{r})
	=
	\int_V
	\bs{j}(\bs{r}')
	\times
	\frac{\bs{r}-\bs{r}'}
	    {|\bs{r}-\bs{r}'|^3}
			d\bs{r}',
	\label{eq:BS}
\end{align}
where $V$ specifies the volume of the disk.

\begin{figure}[b]
\includegraphics[width=0.48\textwidth]{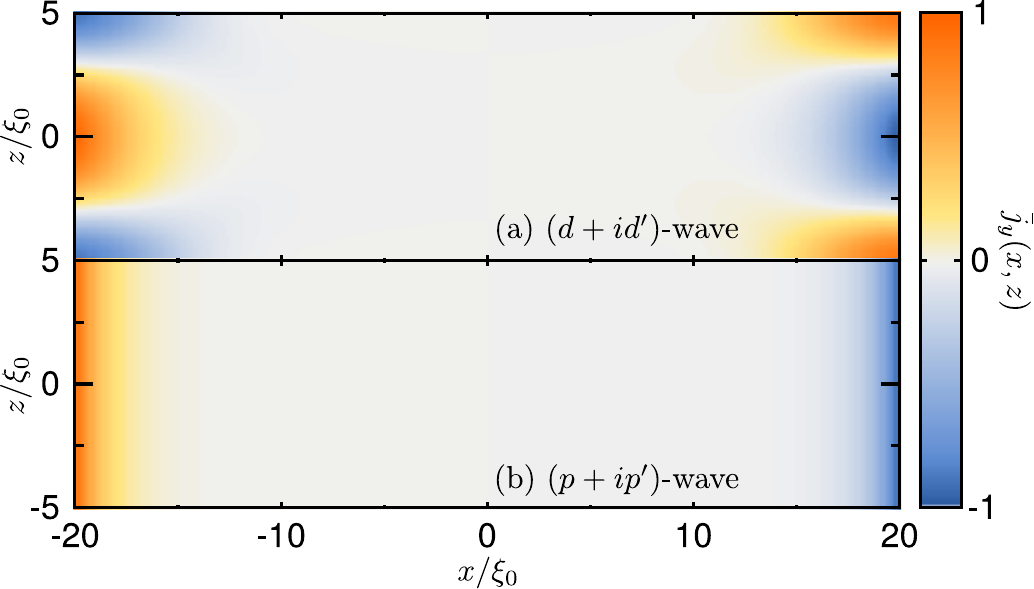}
\caption{Current densities for \textbf{(a)} $(d+id')$- and
	\textbf{(b)} $(p+ip')$-wave SCs.  In the $(d+id')$-wave SC, the
	chiral current is inverted at the top and bottom surface. The
	parameters are $\rho_0=20\xi_0$, $z_0=5\xi_0$, $T=0.2T_{c0}$. 
}
\label{fig:cur}
\end{figure}

\textit{Chiral counter current.}--The current densities for the
($d+id'$)- and ($p+ip'$)-wave SCs are shown in
Fig.~\ref{fig:cur}(a,b), where we fix $y=0$ and $T=0.2T_{c0}$. At the
top/bottom edges ($z=\pm z_0$) of the ($d+id'$)-wave SC, there are
chiral currents flowing in the opposite directions. Since the 
CCC does not appear in the ($p+ip'$)-wave SCs, we conclude that the
CSC is inverted by the flat-band ABS staying at the top/bottom
surfaces of a ($d+id'$)-wave SC 
\footnote{At a surface of an SC, there is an interference between the
incoming and outgoing quasiparticles due to the surface reflection.
When $\Delta( k_\perp, k_\parallel)=-\Delta(-k_\perp,
k_\parallel)$, the flat-band ABS appears at $E=0$.}. 

\begin{figure}[t]
\includegraphics[width=0.46\textwidth]{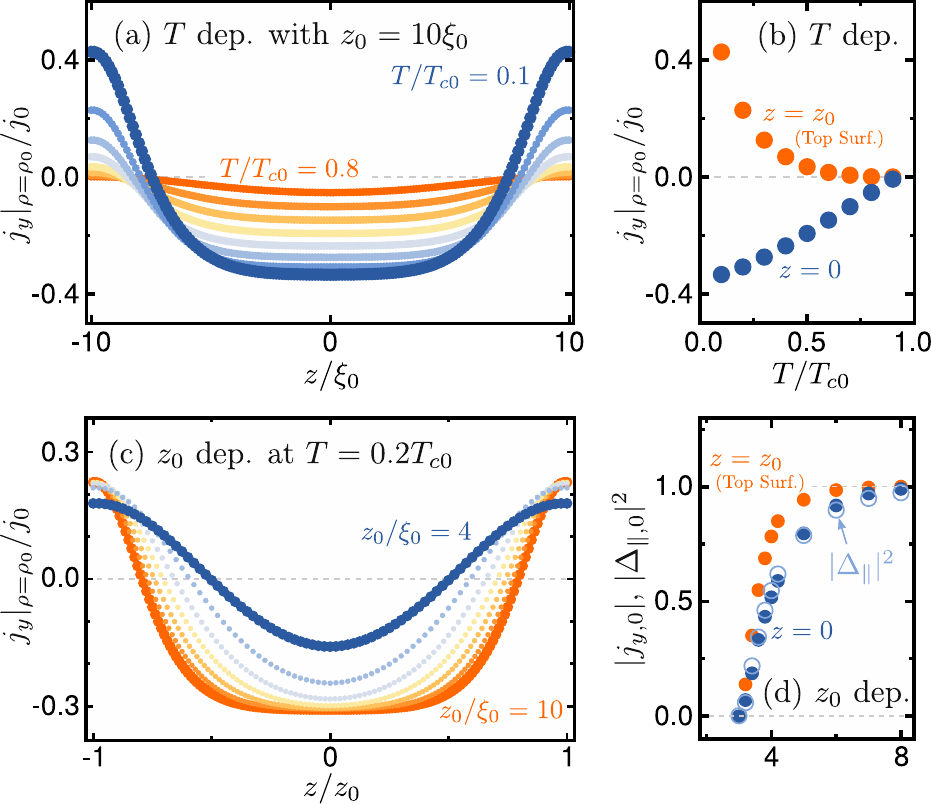}
\caption{
	\textbf{(a,b)} 
	Temperature dependence of the chiral surface current in a
	($d+id$)-wave SC with $z_0=10\xi_0$. The temperature varies from 
	$T/T_{c0}=0.1$ to $0.8$ by $0.1$.
	In \textbf{(b)}, $j_y$ at $z=z_0$ and $0$ are shown.  The CCC is
	enhanced at low temperature. 
	\textbf{(c,d)} 
  Thickness dependence of $j_y|_{\rho=\rho_0}$ with $T=0.2T_{c0}$. 
  In \textbf{(d)}, $|j_y|$ is normalised to
	$|j_y(\rho=\rho_0,z=0)|$ at $z_0=10\xi_0$ and $\Delta_\parallel =
	(\Delta_+-\Delta_-)/2$ is also normalized to its value at
	$z_0=10\xi_0$. 
	The other parameters are the same as those used in Fig.~\ref{fig:cur}. 
}
\label{fig:cur_Tdep}
\end{figure}

The spatial profiles of the current density at $\rho=\rho_c$ for the
\dw are shown in Fig.~\ref{fig:cur_Tdep}(\textbf{a}).  The CSC flows
far from the top/bottom surfaces. The current density at the
lateral surface is strongly influenced near the top/bottom edges. This influence spreads to about $5\xi_0$ from
the top/bottom surfaces.  The current density decreases in amplitude
when approaching the top/bottom surface and changes sign around 
$z_0 -z \sim 3 \xi_0$. 
The amplitudes of the CSC and CCC increase monotonically as $T$
decreases.  However, the $T$ dependences of the respective chiral
currents differ from each other as shown in
Fig.~\ref{fig:cur_Tdep}(\textbf{b}), where $j_y(\rho_c, z=0)$ and
$j_y(\rho_c, z=z_c)$ are shown. 
The CCC exhibits a significant enhancement at low temperatures,
whereas the CSC shows a typical linear behavior at $T\sim T_c$ as
obtained in the \pw SF \cite{Kita_JPSJ_1998}, where $T_c$ is the
critical temperature in the disk geometry. Similar low-temperature
anomalies of the supercurrent and magnetization occur in SCs hosting
flat-band ABSs \cite{Higashitani_PRB_14, Suzuki_PRB_14, Suzuki_PRB_15,
Suzuki_PRR_21}. The details of the temperature dependence are analyzed
in terms of the dispersion relation of the bound states further below. 

The CSC and CCC show different behaviour with respect to the disk
thickness $z_0$ as well.  The $z_0$ dependence of the current density
is shown in Fig.~\ref{fig:cur_Tdep}(c), where $T=0.2T_{c0}$. 
The CCC is less sensitive to $z_0$ compared to the CSC. For
$z_0=5\xi_0$, the CSC is reduced to about 60\% of the value for
$z_0=10\xi_0$, whereas the CCC is barely affected. 
The absolute values of the CSC and CCC and
$|\Delta_{\parallel}(\rho_c,0)|^2$ 
are plotted as a function of $z_0$ in Fig.~\ref{fig:cur_Tdep}(d),
where $\Delta_{\parallel} = (\Delta_+ - \Delta_-)/2$ being the
$d_{yz}$-wave component, $|j_y|$ and $|\Delta_{\parallel}|^2$
 are normalized to their values at $\bs{r}=(\rho_0,0)$ with $z_0=10 \xi_0$. 
The amplitude of the CSC scales well with
$|\Delta_{\parallel}(\bs{r})|^2$, similar to the ($p+ip'$)-wave SF.
Their values are suppressed when $z_0 < 10\xi_0$.  Reducing the
thickness results in the suppression of chiral surface states and of
$\Delta$ by the pair-breaking effect caused by the remaining flat-band ABSs 
at the top/bottom surfaces \cite{HaraNagai_1986,Nagato_PRB_1995}.
Superconductivity breaks down below $z_0 \approx 3.5\xi_0$ in this
configuration.  On the contrary, the CCC is hardly affected by the
thickness when $z_0 > 6\xi_0$.

In a disk geometry, the current-carrying chiral states at the lateral
surface are destroyed by the coupling with the flat-band ABSs. 
This destructive effect is stronger for the chiral states at the
mid-lateral surface (e.g., $z=0$) than for those near 
the top/bottom edge (i.e., $|z|=z_0$). Due to the ballistic motion, the
chiral states at the \textit{top (bottom)} edge are destroyed by the
coupling with the ABSs at the \textit{bottom (top)} surface. These two
states are separated by at least $2z_0$ in real space. On the other
hand, the lateral-surface states [e.g., at $(\rho_0, 0)$] are coupled
with both of the top- and bottom-surface states, only a distance $\sim
z_0$ away. Therefore, the chiral states at the top/bottom edges are
more robust against the thickness reduction than the chiral states at
the lateral surface. The details of the coupling are discussed in the
Supplementary Material \cite{Sup}. 

\begin{figure*}[t!]
\includegraphics[width=0.99\textwidth]{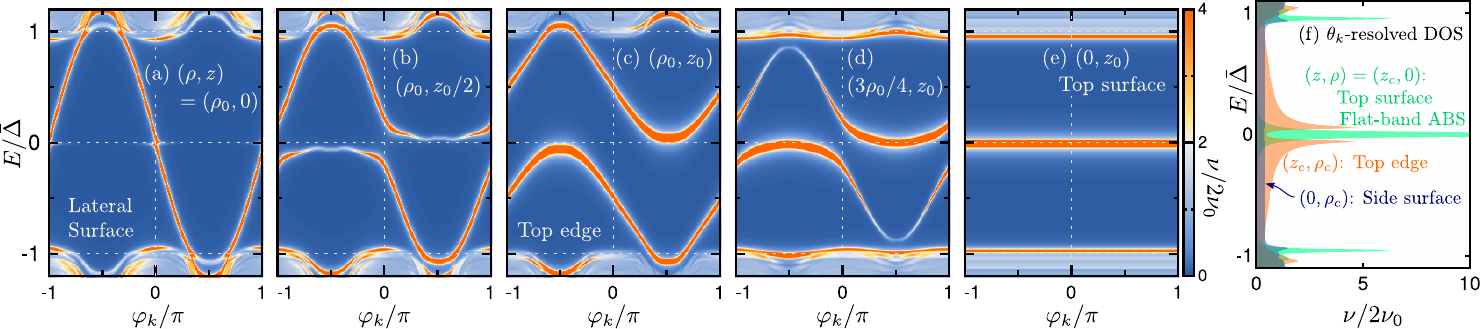}
\caption{
\textbf{(a-e)} Variation of ARDOS. The ARDOS is obtained at 
\textbf{(a)} $(\rho,z) = (\rho_0, 0)$ (i.e., lateral surface), 
\textbf{(b)} $(\rho_0, z_0/2)$, 
\textbf{(c)} $(\rho_0, z_0)$ (i.e., top edge), 
\textbf{(d)} $(3\rho_0/4, z_0)$, and 
\textbf{(e)} $(        0, z_0)$ (i.e., top surface). The ARDOS in (a)
and (e) display the linear dispersion of the chiral surface state and 
flat-band ABS, respectively. 
\textbf{(e)} $\theta_k$-resolved DOS [i.e., $\nu(\bs{r}, \theta_k, E) = \int
\nu(\bs{r}, \vphi_k, k_z, E)$] at the lateral surface, top edge, and top
The parameters are $T=0.2T_{c0}$, $\delta=0.01 \bar{\Delta}(T)$ 
and $\theta_k = \pi/4$. 
}
\label{fig:ARDOS_main}
\end{figure*}

\textit{Angle-resolved density of states.}--The evolution of the ARDOS along the surface is shown in
Fig.~\ref{fig:ARDOS_main}(a-e), where $T=0.2T_{c0}$ and $\theta_k = \pi/4$. 
Note that the ARDOS significantly depends on the spatial dependence of
$\Delta(\bs{r})$, meaning that the self-consistency is
important in analyzing the chiral currents \cite{Sup}. 
At the lateral surface [Fig.~\ref{fig:ARDOS_main}(a)], the ARDOS clearly
shows the chiral surface states as seen in the ($p+ip'$)-wave SC.
At the top surface [Fig.~\ref{fig:ARDOS_main}(e)], a flat-band ABS
appears.
At the top edge, the interaction between the chiral surface state and
the flat-band ABS drastically changes the quasiparticle spectrum as
shown in Fig.~\ref{fig:ARDOS_main}(c): the edge bound state does not
connect the positive- and negative-energy domain but is separated at
$E=0$. 

The ARDOS at the top edge explains the current inversion and its
temperature dependence. At the lateral surface, all of the bound
states are situated in the domains with $E \sin \vphi_k <0$. Those
states carry the charge current in the $-y$ direction (see
Eq.~\eqref{eq:cur_exp2}). 
At the top edge, the bound states appear in all domains in
$E$-$\vphi_k$ space. The contribution from the bound states with $|E|
< \bar{\Delta}_\theta/2$ carry a positive current and are larger than
that of the $|E| \geq \bar{\Delta}_\theta/2$ region, where
$\bar{\Delta}_\theta = \bar{\Delta} \cos \theta_k$.  The ARDOS for the
edge bound states is maximal around $(E, \vphi_k)=(0, \pm \pi/2)$ and
decreases monotonically as $|E|$ increases. 

The $\theta_k$-resolved DOS $\nu(\theta_k)$ is
shown in Fig.~\ref{fig:ARDOS_main}(f), where $\nu(\theta_k)$ is
calculated from the ARDOS in Fig.~\ref{fig:ARDOS_main}(a,c,e). The DOS
for the chiral surface states is flat for $|E|<\bar{\Delta}_\theta$, as
obtained in the $(p+ip')$-wave case. At the top surface, $\nu(\theta_k)$
exhibits a zero-energy peak reflecting the flat-band ABSs. 

At the edge, two narrowly split peaks appear at $E \approx 0.06
\bar{\Delta}$ in $\nu(\theta_k)$. We have confirmed the split width
decreases with increasing $\rho_0$.
Comparing Figs.~\ref{fig:ARDOS_main}(c) and \ref{fig:ARDOS_main}(f),
it is seen that the split peaks consist of the bound
states carrying the CCC.  Those low-energy peak
cannot contribute to $j_y$ at high temperatures $T \sim T_{c0}$
because of the factor $\tanh ( {E}/{2T} ) $ coming from the
Fermi--Dirac distribution function. On the contrary, at low
temperatures, the contribution of the low-energy states increases
rapidly. 

\begin{figure}[b]
\centering
\includegraphics[width=0.46\textwidth]{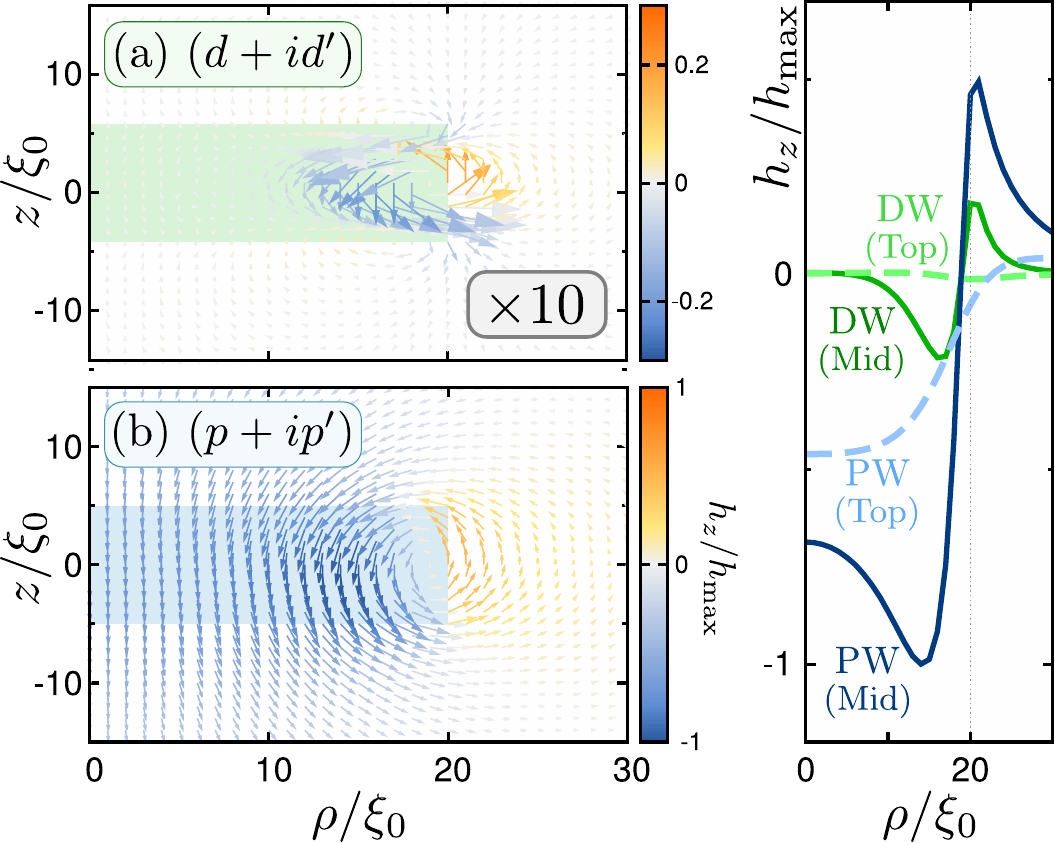}
\caption{ Vector plots of $\bs{h}(\bs{r})$ for \textbf{(a)} ($d+id'$)-
and \textbf{(b)} ($p+ip'$)-wave SCs. 
The arrows in (a) are magnified by a factor of 10 for better 
visibility.
\textbf{(c)} Profile $h_z(\bs{r})$ at $z=0$ (middle) and $z=z_c$
(top surface). The magnetic field is normalized to its maximum
amplitude in the \pw case. The parameters
are: $\rho_0=20\xi_0$, $z_0=5\xi_0$, and $T=0.2T_{c0}$. }
\label{fig:field}
\end{figure}

The details of the temperature dependence can be analytically examined
by modeling the bound-state quasiparticle spectra. We model the
quasiparticle spectrum for the chiral edge states in
Fig.~\ref{fig:ARDOS_main}(c) as 
\begin{align}
	& \nu_\vphi(E,\theta_k) = \sum_{\pm} 
	A_\vphi^\pm \delta(E-E_\vphi^\pm), 
	\\
  & E_\vphi^\pm = \pm \frac{\Delta}{\pi^2} 
	\left( \vphi \mp \frac{\pi}{2} \right)^2, 
	\\
  & A_\vphi^\pm = \Delta - \frac{\Delta}{\pi^2}
	\left( \vphi \mp \frac{\pi}{2} \right)^2, 
  \label{eq:spec_main}
\end{align}
where $|\vphi_k| \leq \pi/2$ and the $\theta_k$-resolved DOS satisfies
$\nu_\vphi(E,\theta_k) = \nu_{\vphi+\pi}(E,\pi-\theta_k)$
\footnote{This model is justified in the $\rho_0 \to \infty$ limit
where $\nu(\bs{r},E)$ does not show the energy splitting at $E=0$.}.
This form is motivated by the numerical results in
Figs.~\ref{fig:ARDOS_main}(c).  Using Eq.~\eqref{eq:cur_exp2} and a
linear interpolation of the  distribution function, we obtain
analytical expressions for the corresponding temperature dependence of
the chiral current density \cite{Sup}. 
The CCC behaves as $j_{T}-j_{T=0} \sim -\sqrt{T/\bar{\Delta}}$ at $T
\ll \bar{\Delta}$, whereas the CSC behaves as $j_{T}-j_{T=0}  \sim
(T/\bar{\Delta})^2$, as demonstrated in 
Refs.~\cite{Sauls_PRB_2011, Tsutsumi_PRB_12, Tsutsumi_JPSJ_12}, where
$j_T$ is the $y$-component of the $\theta_k$-resolved current density
at $T$.  This temperature dependence is consistent with the
low-temperature enhancement in Fig.~\ref{fig:cur_Tdep}(b). 

Using the Biot-Savart law, the local magnetic field $\bs{h}(\bs{r})$
is obtained from $\bs{j}(\bs{r})$. The vector
plots of $\bs{h}(\bs{r})$ obtained from Fig.~\ref{fig:cur} are shown
in Fig.~\ref{fig:field}(a,b), where $h_z$ is normalized to the maximum
value of $|\bs{h}|$ in the \pw SC. 
In a \pw SC [Fig.~\ref{fig:field}(b)], the spontaneous magnetic field
is sufficiently large to be detected in experiment
\cite{Matsumoto_JPSJ_1999, Moler_05, Nelson_07}. On the other hand,
the spontaneous magnetic field for the \dw is much smaller than that
in the \pw case. The CCCs at the top/bottom edges reduce the effect of
the lateral chiral currents on $\bs{h}$. 
For example, $\bs{h}$ at the top surface is significantly reduced by
the CCC at the top edge compared to the case of a $p+ip'$
superconductor. 
The $z$-component of $\bs{h}(\bs{r})$ at the middle ($z=0$) and the
top ($z=z_0$) of the disk are shown in Fig.~\ref{fig:field}(c). The
results for a thicker disk is shown in Supplementary Material \cite{Sup}. Note that the
scanning SQUID measurements that detect the spontaneous magnetization
are typically performed above the top surface of a sample. In the \pw case, the
magnetic field is sufficiently large at the top surface to be detected by SQUID measurements. At the
top surface of the \dw SC, the magnetic field is almost zero. We have
confirmed that the maximum amplitude of $|h_z(z=z_0)|$ in the \dw SC
is about 4\% of that	in the \pw. 

\textit{Conclusions.}--We have analyzed the spontaneous chiral surface current
in a disk-shaped $(d_{zx}+id_{yz})$-wave superconductor. We have demonstrated that
the chiral current is reversed at the top/bottom edges of the disk as a result of
 flat-band Andreev bound states situated at the top/bottom surface,
and that the net chiral current is significantly reduced compared to the
$(p_x+ip_y)$-wave case. Accordingly, the spontaneous
magnetic field generated by the chiral currents, observable in
experiment, is significantly suppressed for a $(d_{zx}+id_{yz})$-wave
superconductor, rendering such a superconductor an excellent candidate
for chiral superconductivity in systems lacking the observation of a
spontaneous magnetization.

\begin{acknowledgments}
We are grateful to Y.~Asano for the fruitful
discussions.  S.-I.~S. is supported by JSPS Postdoctoral Fellowship
for Overseas Researchers, and would like to thank the University of
Twente for hospitality.
\end{acknowledgments}

\pagebreak
\setcounter{equation}{0}
\setcounter{figure}{0}
\setcounter{section}{0}
\setcounter{page}{0}
\makeatletter
\renewcommand{\thesection}{S\arabic{section}}
\renewcommand{\theequation}{S\arabic{equation}}
\renewcommand{\thefigure}{S\arabic{figure}}
\renewcommand{\bibnumfmt}[1]{[S#1]}
\renewcommand{\citenumfont}[1]{S#1}

\clearpage
\onecolumngrid
\begin{center}
\textbf{\large Supplemental Materials: } \\ 
\vspace{.1cm}
\textbf{\large Chiral Current Inversion by Flat-Band Andreev Bound States}\\
\vspace{.5cm}
Shu-Ichiro  Suzuki$^{1}$, 
Alexander A. Golubov$^{1}$, and  M. Eschrig$^{2}$\\
\vspace{.1cm}
\small{ 
$^{1}$\it{MESA+ Institute for Nanotechnology, University of Twente, 
7500 AE Enschede, The Netherlands} \\
$^{2}$\it{University of Greifswald, Institute of Physics, 17489 Greifswald, Germany} 
}
\end{center}

\section{Temperature dependence of the chiral currents}
\subsection{Chiral surface current at lateral surface}

The current density is given by 
\begin{align}
  \bs{j} 
	& = -\frac{e v_F \nu_0 \pi}{4} 
	\int 
	\langle \bs{k} \tr{\check{\tau}_3 \check{g}^K} \rangle
	\frac{dE}{2\pi},
	\\
	& = |e| v_F 
	\int_{-\infty}^{\infty}
	\left \langle \bs{k} \nu_{\bs{k}} 
	\right \rangle \tanh \left( \frac{E}{2T} \right) 
	dE,
	\\
	\nu_{\bs{k}} &=  \nu_0({g}^R-{g}^A) = 2  \nu_0\re{g^R}
\label{}
\end{align}
where for an equilibrium system $\check{g}^K =
(\check{g}^R-\check{g}^A) f$ with $f(E,T) = \tanh(E/2T)$ being the
distribution function, $\nu_{\bs{k}}$ the angle-resolved local
density of states (ARDOS), $e<0$ the charge of an quasiparticle,
$v_F$ the Fermi velocity, and $\nu_0$ the density of states (DOS)
per spin at the Fermi level. The current density along the lateral
surface is 
\begin{align}
  & {j_y}(\theta)
	= 
	\frac{j_0}{2}\sum_{\pm k_z}
	\int_{-\infty}^{\infty}
	\int_{-\pi}^{\pi}
	\nu_{\vphi,k_z} \sin \theta \sin \vphi 
	f(E, T)
	\frac{d \vphi}{2 \pi }dE,
	\\
	%
	%
  & J = \frac{j_y(\theta)}{j_0 \sin \theta} = 
	\int_{-\infty}^{\infty}
	\int_{-\pi/2}^{\pi/2}
	\nu_{\vphi} \sin \vphi 
	{f}
	\frac{d \vphi}{\pi }dE,
  \label{eq:cur_den_ana}
\end{align}
where $J=J(\theta)$ is the $\theta$-resolved normalized current density and 
$j_0=|e| v_F \nu_0$. The direction of the momentum is given by 
$\bs{k}= (
\sin \theta \cos \vphi, 
\sin \theta \sin \vphi, 
\cos \theta)$. 
In eq.~\eqref{eq:cur_den_ana}, we have used the symmetry of the
quasiclassical Green's function (QGF) and omitted the subscript $k_z$.
The boundary condition $\check{g}^R(\bs{r}_{\text{b}},
\bs{k}_{\parallel}, k_{\perp}, E) =\check{g}^R(\bs{r}_{\text{b}},
\bs{k}_{\parallel},-k_{\perp}, E)$ gives us $\nu_{\vphi,k_z} =
\nu_{\pi-\vphi,k_z}$ at the lateral surface, where
$\bs{k}_{\parallel}$ and $k_{\perp}$ are the parallel and
perpendicular components of the momentum respectively and
$\bs{r}_{\text{b}}$ represents the boundary. At a edge parallel to the
$y$ direction, the symmetry $\nu_{\vphi,k_z} = \nu_{\vphi,-k_z}$ is
also hold. For the chiral lateral-surface state, $\nu_{\vphi,k_z} =
\nu_{\vphi,-k_z}$ is hold under $\Delta(x,y,z) = \Delta(x,y,-z)$. 


The quasiparticle spectrum of the surface bound states can be modeled
by the Dirac's delta function.  The spectrum for the lateral-surface
states of a chiral SC can be modeled as 
\begin{align}
  & \nu_\vphi(E) =  A_\vphi \delta(E-E_\vphi), \\
  & E_\vphi = - \Delta \sin \vphi, 
	\hspace{6mm}
	A_\vphi
	= \Delta \cos \vphi, 
  \label{eq:spec-p}
\end{align}
where $\Delta = \Delta(\theta)$ is the $\theta$-dependent pair pair
potential. This spectral function is estimated from the
self-consistent solution with $\Delta_{\bs{k}} = \Delta_0[k_x
\tanh(x/\xi_0) + i k_y]$ (see [Y.~Tsutsumi and K.~Machida,
Phys.~Rev.~B \textbf{85}, 100506(R) (2012).
J.~Phys.~Soc.~Jpn.~\textbf{81}, 074607 (2012)] for the details). 

The current density becomes
\begin{align}
  J & =
	\int_{-\pi/2}^{\pi/2}
	\sin \vphi \cos \vphi
	{f}(E_\vphi)
	\frac{d \vphi}{\pi }
	%
  =
	2\int_{0}^{\pi/2}
	\sin \vphi \cos \vphi
	f(E_\vphi)
	\frac{d \vphi}{\pi },
	\\
  & \approx
	\frac{2 \Delta}{\pi} \int_{0}^{\pi/2}
	\sin \vphi \cos \vphi
	f'(E_\vphi)
	d \vphi,
\end{align}
where we have introduced the approximated distribution
functions in order to obtain the analytic form: 
\begin{align}
  f'(E) = \left\{ \begin{array}{cc}
	  E/2T          & \text{for $|E|=2T$, } \\
		\mathrm{sgn}[E] & \text{otherwise.}  \\
  \end{array} \right.
\end{align}
The current density in each region can be calculated as 
\begin{align}
  J & =
	- \frac{2\Delta}{\pi}
	\left\{
	\frac{1}{\alpha}
	\int_{0}^{\vphi_0}
	\sin^2 \vphi \cos \vphi
	d \vphi
	+
	\int_{\vphi_0}^{\pi/2}
	\sin \vphi \cos \vphi
	d \vphi
	\right\}, 
	\\
  & =
	- \frac{2\Delta}{\pi}
	\left\{
	\frac{1}{3 \alpha}
	\left[ \sin^3 \vphi \right]_{0}^{\vphi_0}
	+ \frac{1}{2}
	\left[ \sin^2 \vphi \right]_{\vphi_0}^{\pi/2}
	\right\}, 
	\\
  & =
	- \frac{2\Delta}{\pi}
	\left\{
	\frac{\alpha^2}{3 }
	+ \frac{1}{2}
	\left[ 1 - \alpha^2 \right]
	\right\} 
	= 
	- \frac{\Delta}{\pi}\left(1-\frac{\alpha^2}{3}\right), 
\end{align}
where $2T = E_{\vphi_0}$ and $\alpha = \sin \vphi_0 = 2T/\Delta$.
Therefore, the temperature dependence of the $\theta$-resolved current
density is $J(T)-J(T=0) \sim T^2$. This result coincides with the
results in previous work.

\begin{figure*}[tb]
\centering
\includegraphics[width=0.6\textwidth]{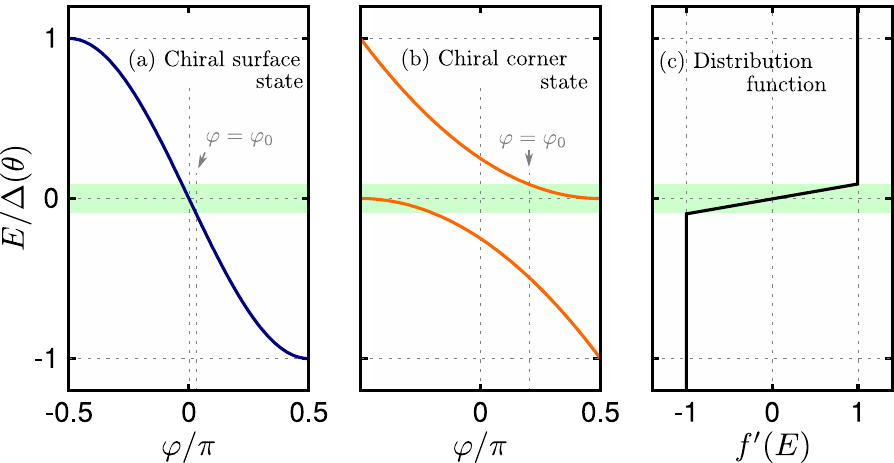}
\caption{Modeled dispersion relations of the \textbf{(a)} chiral surface state and
\textbf{(b)} the edge state. \textbf{(c)} Approximated distribution
function. The colored regions indicate the region where the
distribution function $f'$ is linear. }
\label{fig:disp_analy}
\end{figure*}
\subsection{Chiral counter current at the edge}
From the numerical results, we model the spectrum for
the edge states of the ($d+id'$)-wave superconductor (SC) as 
\begin{align}
  & \nu_\vphi(E) = \sum_{\pm} A_\vphi^\pm \delta(E-E_\vphi^\pm), \\
  & E_\vphi^\pm = \pm \frac{\Delta}{\pi^2} 
	\left( \vphi \mp \frac{\pi}{2} \right)^2, 
	\hspace{6mm}
    A_\vphi^\pm = \Delta - \frac{\Delta}{\pi^2}
		\left( \vphi \mp \frac{\pi}{2} \right)^2. 
\label{eq:spec}
\end{align}
The weight of the spectrum $A_\vphi^\pm$ is estimated 
using the numerical results. The functions in Eq.~\eqref{eq:spec} have
the symmetry: $E_{-\vphi}^\pm = - E_\vphi^\mp$ and $A_{-\vphi}^\pm =
A_\vphi^\mp$. Using these functions, the current density is reduced to
\begin{align}
  J   
	& =
	\sum_\pm
	\int_{-\pi/2}^{\pi/2}
	\sin \vphi A^\pm_\vphi
	f(E^\pm_\vphi)
	\frac{d \vphi}{\pi },
	\\
	& =
	2 \int_{-\pi/2}^{\pi/2}
	\sin \vphi A^+_\vphi
	f(E^+_\vphi)
	\frac{d \vphi}{\pi }. 
\end{align}
Hereafter, we will omit the superscript $\pm$. 

The current density can be separated into two contributions: $J_{\mathrm{I}}$ 
from $|E|<2T$ and $J_{\mathrm{II}}$
from $|E| \geq 2T$. 
The first contribution can be calculated as 
\begin{align}
  J_{\mathrm{I}}
	& =
	\frac{2 \Delta}{\pi \alpha_0^2}
	\int_{\vphi_0}^{\pi/2}
	\sin \vphi 
	\left[ 1 - \frac{1}{\pi^2}
		\left( \vphi - \frac{\pi}{2} \right)^2\right]
    \left( \vphi - \frac{\pi}{2} \right)^2
	d \vphi, 
	\\
	& =
	\frac{2 \Delta}{\pi \alpha_0^2}
	\int_{0}^{\alpha_0}
	X^2 \cos X
	\left[ 1 - \frac{X^2}{\pi^2}\right]
	dX ,
	\\
	& =  
	\frac{2 \Delta}{\pi \alpha_0^2}
	\left[ C_2 - \frac{1}{\pi^2} C_4 \right],
	\label{eq:tmp01}
\end{align}
where 
the angle $\vphi_0$ is defined by the relation: $2T = E^+_{\vphi_0}$.
For the chiral edge states, $\vphi_0$ is given by $
	\vphi_0 = \pi/2 - \alpha _0
	$ 
	with the parameter
\begin{align}
	\alpha_0 = \sqrt{{2 T \pi^2}/{\Delta}}
\end{align}
being extensively used below quantifying the temperature dependence.
The coefficients $C_n = C_n(A, B)$ is defined by $C_n \equiv \int_A^B
X^n \cos X dX$. Using Eqs.~\eqref{eq:cos1}-\eqref{eq:cos3} written
below, we have the temperature dependence 
\begin{align}
  J_{\mathrm{I}}
	& =
	\frac{2 \Delta}{\pi^3}
	\left\{
	(\pi^2 + 8) \left[\sin \alpha_0 -\frac{2}{\alpha_0}
	\left(  \frac{\sin \alpha_0}{\alpha_0} - \cos \alpha_0 \right)
	\right]
  -4 \alpha_0^2 \sin \alpha_0 - 3 \alpha_0 \cos \alpha_0
	\right\}. 
	\label{eq:tmp02}
\end{align}
At the low-temperature limit (i.e., $\alpha \ll 1$), the contribution
$J_I$ becomes zero. 
Note the limit of the sinc function can be obtained by the
L'H\^{o}pital's rule: 
\begin{align}
  \lim_{x \to 0}~ \left\{
	\sin x -
	\frac{2}{x} 
	\left( \frac{\sin x}{x} - \cos{x} \right)
	\right\}
	=
  \lim_{x \to 0}~ 
	\frac{x \cos x}{2}
	= x . 
\end{align}
The second contribution can also be calculated, 
\begin{align}
  J_{\mathrm{II}}
	& =
	\frac{2 \Delta}{\pi}
	\int_{\alpha_0}^{\pi}
	\cos X
	\left[ 1 - \frac{X^2}{\pi^2}\right]
	dX 
	=  
	\frac{2 \Delta}{\pi}
	\left[ C_0(\alpha_0, \pi) - \frac{1}{\pi^2} C_2(\alpha_0, \pi) \right],
	\\
	& =  
	\frac{2 \Delta}{\pi^3}
	\left[ 
    2\pi - 2 \alpha_0 \left(\frac{\sin \alpha_0}{\alpha_0}-\cos \alpha_0 \right)
		+(\alpha_0^2-\pi^2) \sin \alpha_0
	\right]. 
	\label{eq:tmp03}
\end{align}
At zero temperature, $J_{\mathrm{II}} = J(T=0) = {4 \Delta}/{\pi^2}$. 
The total current $J(T)$ is given by 
\begin{align}
  J(T) = 
	\frac{2 \Delta}{\pi^3}
	\left[ 
	2 \pi+
	6 \sin  \alpha_0
	-\alpha_0 \cos \alpha_0
	-\frac{\pi^2+8}{\alpha_0}
	\left(\frac{\sin \alpha_0}{\alpha_0}-\cos \alpha_0 \right)
	-3 \alpha_0^3 \sin \alpha_0
	\right]. 
\label{}
\end{align}
In the low-temperature limit, the total current becomes 
\begin{align}
  J(T) \to 
	\frac{2 \Delta}{\pi^3}
	\left[ 
	2 \pi
	-(\pi^2+8)\alpha_0 
	\right].
\label{}
\end{align}
Consequently, the total current for the chiral counter current behaves
as $J(T) - J(T=0) \sim \sqrt{T/\Delta}$. 

The coefficients $C_n=C_n(A, B)$ above have been evaluated as follows: 
\begin{align}
  C_0 
	& = \int_A^B \cos X dX
	= \left[ \, \sin X \right]_A^B, 
	\label{eq:cos1}
	\\[2mm]
  C_2
	& =  \left[ \, X^2 \sin X  + 2X \cos X \right]_A^B-2C_0, \\
	& = \left[ \, (X^2-2) \sin X  + 2X \cos X  \right]_A^B, 
	\\[2mm]
  C_4
	& =  \left[ \, X^4 \sin X  + 3X^3 \cos X \right]_A^B-8C_2, \\
	& = \left[ \, (X^4-8X^2+16) \sin X  + 4X (X^2-4) \cos X \right]_A^B. 
	\label{eq:cos3}
\end{align}

\begin{figure*}[bt]
\centering
\includegraphics[width=0.96\textwidth]{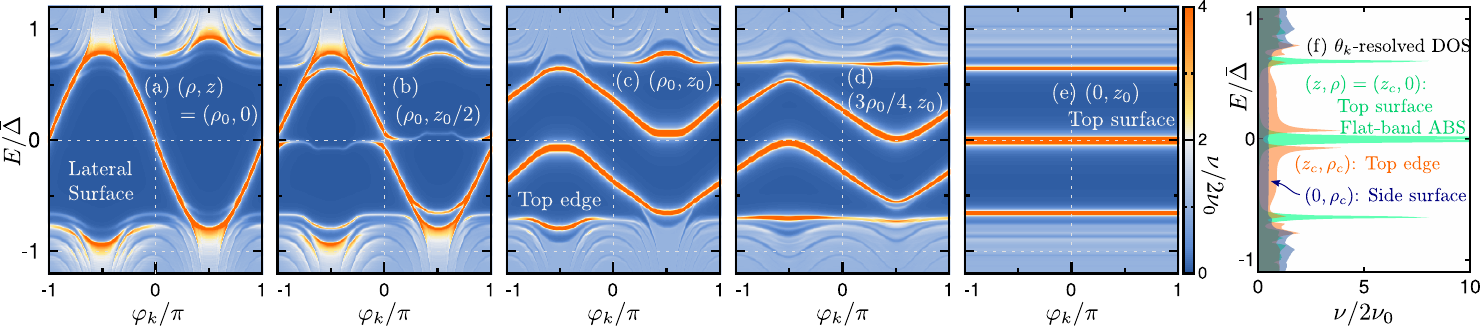}
\caption{
Angle-resolved density of states with $\theta_k = 3\pi/8$.
{\textbf{(a-e)} Variation of ARDOS. The ARDOS is obtained at 
\textbf{(a)} $(\rho,z) = (\rho_0, 0)$ (i.e., lateral surface), 
\textbf{(b)} $(\rho_0, z_0/2)$, 
\textbf{(c)}$(\rho_0, z_0)$ (i.e., top edge), 
\textbf{(d)} $(3\rho_0/4, z_0)$, and 
\textbf{(e)} $(        0, z_0)$ (i.e., top surface). The ARDOS in (a)
and (e) display the linear dispersion of the chiral surface state and 
flat-band ABS, respectively. 
\textbf{(e)} $\theta_k$-resolved DOS [i.e., $\nu(\bs{r}, k_z, E) = \int
\nu(\bs{r}, \vphi_k, k_z, E)$] at the lateral surface, top edge, and top
surface. The top-edge state shows a peak that give a more contribution
at low temperatures. 
The parameters are $T=0.2T_{c0}$, $\delta=0.01 \bar{\Delta}(T)$, and $k_z = 1/\sqrt{2}$. 
}}
\label{fig:dos_FR_3-8_T20}
\end{figure*}

\begin{figure*}[bt]
\centering
\includegraphics[width=0.96\textwidth]{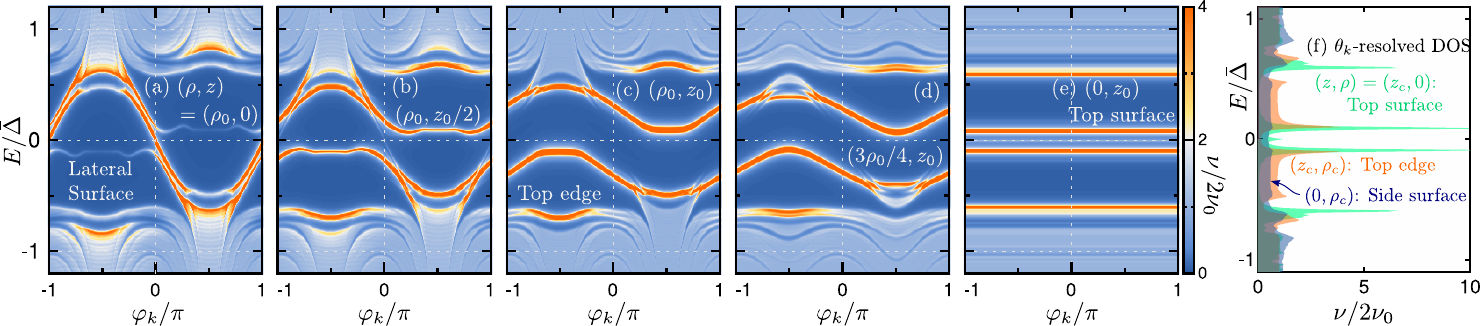}
\caption{
Angle-resolved density of states with $\theta_k =
3\pi/8$. The results are plotted in the same manner as in
Fig.~\ref{fig:dos_FR_3-8_T20} but are calculated at $T=0.8T_{c0}$. 
The ZEP at the top surface splits because of 
the coupling with the chiral surface states at the lateral surface. 
}
\label{fig:dos_FR_3-8_T70}
\end{figure*}

\begin{figure*}[tb]
\centering
\includegraphics[width=0.90\textwidth]{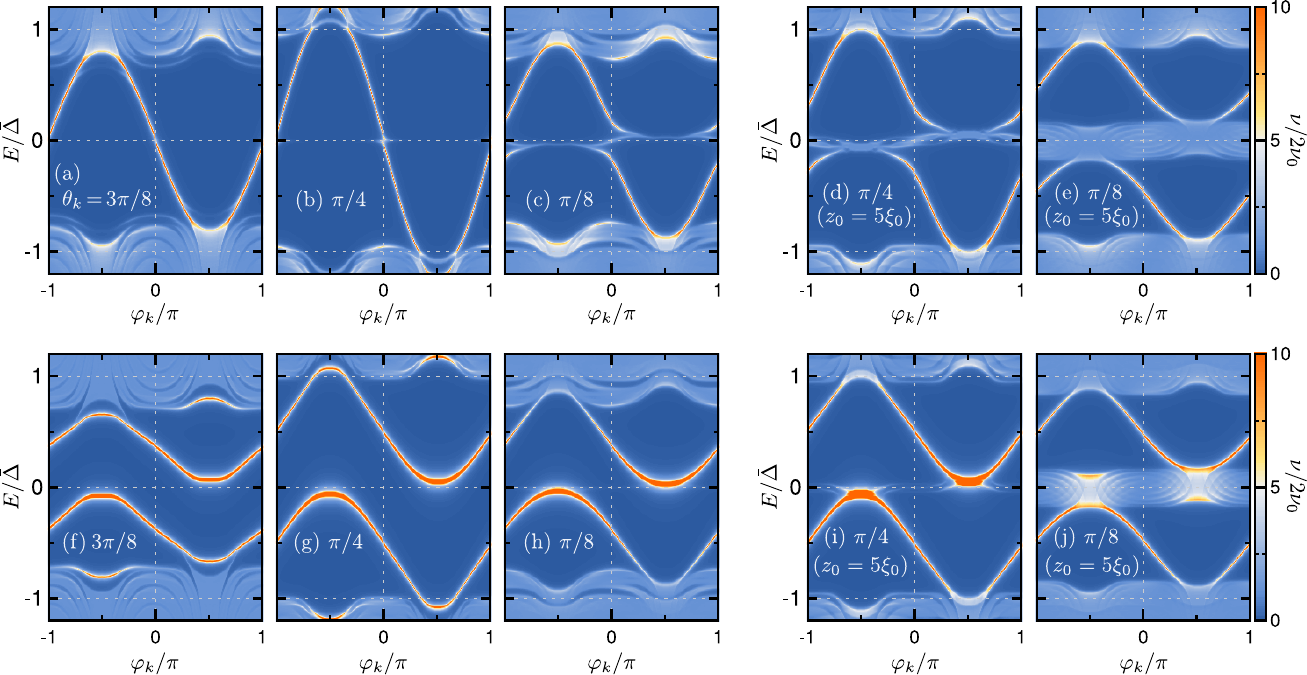}
\caption{
Angle-resolved density of states with different $\theta_k$. 
	The results are obtained at \textbf{(a-e)} lateral-surface and \textbf{(f-j)} top-edge states. 
  The angle is set to 
	(a,f) $\theta_k=3\pi/8$, 
	(b,d,g,i) $\theta_k=\pi/4$, and 
	(c,e,h,j) $\theta_k=\pi/8$. 
	The thickness of the disk is $z_0=10\xi_0$ in the left panels
	(a, b, c, f, g, h) and $z_0=5\xi_0$ in the right panels (d, e, i, j). 
	When the disk thickness is $z_0=5\xi_0$, the multiple reflections at
	the top/bottom surfaces generate new low-energy states. 
	The parameters are $T=0.2T_{c0}$ and $\delta=0.01 \bar{\Delta}(T)$. }
\label{fig:ARDOS_theta}
\end{figure*}

\section{Angle-resolved density of states}
In this section, we discuss the $\theta_k$-dependence of the ARDOS. 
The ARDOS are shown in Figs.~\ref{fig:dos_FR_3-8_T20} and \ref{fig:dos_FR_3-8_T70}, where the angle is
fixed at $\theta = 3\pi/8$. The temperature is fixed at $T=0.2T_{c0}$ 
($0.8T_{c0}$) in Fig.~\ref{fig:dos_FR_3-8_T20} (Fig.~\ref{fig:dos_FR_3-8_T70}). The smearing
factor is $\delta=0.01\bar{\Delta}(T)$. 
When the angle $\theta_k$ is closer to $\pi/2$ ($k_z$ is smaller), the
chiral surface states are less affected by the coupling with the
top/bottom surfaces as shown in 
Figs.~\ref{fig:ARDOS_main}(a) and \ref{fig:dos_FR_3-8_T20}(a). 
When $k_z$ is large, the influence from the
top/bottom surface to the chiral surface states at the lateral surface
becomes large. The dispersion
relation at $E=0$ with $\theta_k = 3\pi/8$ [Fig.~\ref{fig:dos_FR_3-8_T20}(a)] is closer to linear than that with
$\theta_k = \pi/4$ [Fig.~\ref{fig:ARDOS_main}(a)]
[see the discussion related to Fig.~\ref{fig:ARDOS_theta}].
The energy splitting of the top-edge states becomes larger than that for the
$\theta_k = \pi/4$ case [Compare Figs.~\ref{fig:ARDOS_main}(f) and
\ref{fig:dos_FR_3-8_T20}(f)]. In a disk geometry, there are
quasiclassical paths that are constructed along the perimeter. 
These paths undergo multiple reflections, leading to dephasing. 
A quasiclassical path with larger in-plane momentum (i.e.,
$\theta_k \approx \pi/2$) experiences more multiple reflections than one with smaller in-plane momentum. 
Consequently, the
energy splitting for $\theta_k = 3\pi/8$ is larger than that for
$\theta_k = \pi/4$.  

Comparing Figs.~\ref{fig:dos_FR_3-8_T20} and \ref{fig:dos_FR_3-8_T70},
we see that the temperature does not qualitatively modify the ARDOS
except for the zero-energy peak (ZEP). 
The ZEP at the top/bottom surface splits when the temperature
increases [Figs.~\ref{fig:dos_FR_3-8_T70}(e) and
\ref{fig:dos_FR_3-8_T70}(f)].  At a higher temperature, the ZEP
interacts with the one at the opposite surface. The ZEP splits due to
the coupling between those zero-energy states.  With increasing
temperature, the coherence length $\hbar v_F/ \bar{\Delta}(T)$ becomes
longer as the pair potential decreases. 

The ARDOS at $(\rho,z)=(\rho_0,0)$ (i.e., lateral surface) are shown
in Fig.~\ref{fig:ARDOS_theta}(a-e), where the angle is set  to (a)
$\theta_k = 3\pi/8$, (b,d) $\pi/4$, and (c,e) $\pi /8$, the disk
thickness is set to (a-c) $z_0=10\xi_0$ and (d,e) $5\xi_0$.
Figure~\ref{fig:ARDOS_theta}(a-c) demonstrate the effect of the
coupling with the flat-band ABSs. 
The chiral surface states with larger $k_z = \cos \theta_k$ are more
influenced, in particular, at low energies. The influence from the
flat-band ABSs becomes more significant when the disk is thinner
[Fig.~\ref{fig:ARDOS_theta}(d,e)]. 

The chiral edge states, on the other hand, are not qualitatively
modified by varying $\theta_k$.  The ARDOSs at $(\rho,z)=(\rho_0,z_0)$
(i.e., top edge) are shown in Fig.~\ref{fig:ARDOS_theta}(f-j). 
The low-energy states at the \textit{top} edge can be coupled with the
flat-band ABS at the \textit{bottom} surface. The distance between the
top edge and the bottom surface is at least $2z_0$, whereas the
distance between the point $(\rho_0,0)$ and the top/bottom surface is
at least $z_0$. Since the bound states decays exponentially from the
surface, the distance between low-energy states are important. 
Therefore, the edge states are more robust than the lateral-surface
states.  When the disk is thinner ($z_0=5\xi_0$), the influence of the
coupling with the flat-band ABS at the bottom surface appears at low
energies [Fig.~\ref{fig:ARDOS_theta}(i,j)].

The chiral surface state is modified by the flat-band ABSs at the
top/bottom surfaces, whereas the chiral edge state at the top edge is
modified by the flat-band ABSs sitting at the bottom surface, where
the top edge and bottom surface are separated by at least $z_0$. 

When the disk thickness is $z_0=5\xi_0$, the multiple reflections at 
the top/bottom surfaces generate new low-energy states. 

\begin{figure*}[tb]
\centering
\vspace{10mm}
\includegraphics[width=0.98\textwidth]{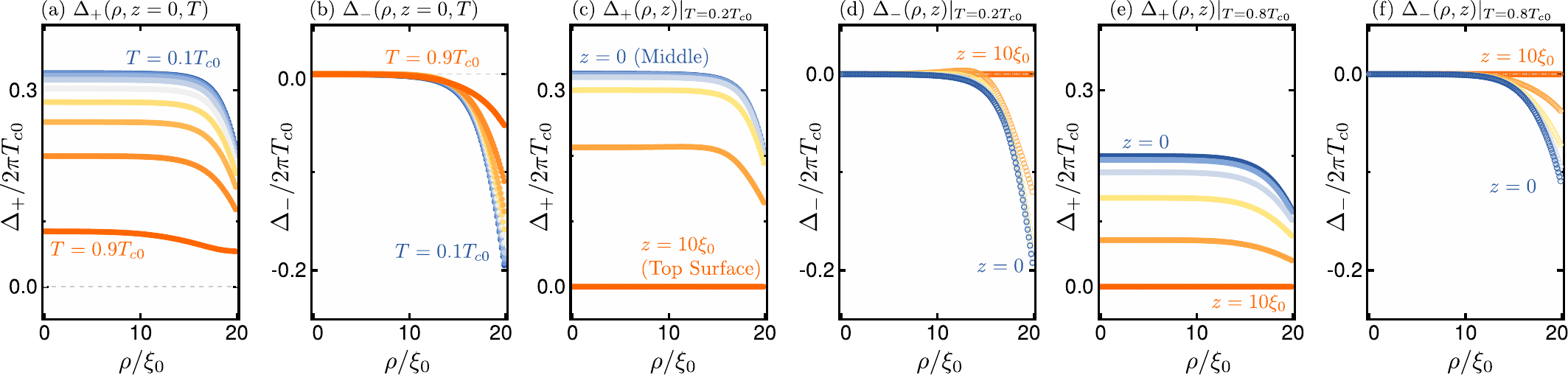}
\caption{(a,b) Temperature dependence of pair potentials at $z=0$ in
($d+id'$)-wave disk. Spatial dependence of pair potentials at 
(c,d) $T=0.2T_{c0}$ and (e,f) $T=0.8T_{c0}$. The dimension of the disk is set
to $\rho_c = 20 \xi_0$ and $z_c = 10 \xi_0$. 
}
\label{fig:del}
\end{figure*}
\section{\textbf{Pair potential}}

The pair potentials for the ($p+ip'$)-wave SC are shown in
Fig.~\ref{fig:del}(a), where the results are normalized to its bulk
value and $T=0.2T_{c0}$. At the surface, both components have the same
amplitude ($\Delta_+ = \Delta_-$) because the $k_x$ component is
completely killed by the specular reflections. The other component, on
the other hand slightly enhanced. This behavior is consistent with
[M.~Matsumoto and and M.~Sigrist, {J. Phys. Soc. Jpn. \textbf{68} 3
(1999)}.  A.~Furusaki, M.~Matsumotoand and M.~Sigrist,
\href{https://link.aps.org/doi/10.1103/PhysRevB.64.054514}{Phys.~Rev.~B
\textbf{64}, 054514 (2001)}]. 

Differing from the ($p+ip'$)-wave case, the pair potentials depend on
$z$ in the ($d+id'$)-wave SC. The pair potentials for the
($d+id'$)-wave SC are shown in Fig.~\ref{fig:del}(b-d). At the middle
of the disk ($z=0$), the pair potentials show the qualitatively the
same behaviors as those in the ($p+ip'$)-wave SC. With approaching to
the top (or bottom) surface, the amplitudes of the pair potentials
decrease. The suppression of $\Delta_\pm$ is caused by the pair
breaking by the ABSs as happens in $d_{xy}$-wave SCs. 

\section{\textbf{Dispersion relation for surface Andreev bound states
under homogeneous pair potential}}

Both of the ($d+id'$)- and ($p+ip'$)-wave SCs host the chiral ABS at
its lateral surface. The dispersion relation for the chiral state can be
obtained analytically under the non-self-consistent pair potential. 

\begin{figure*}[tb]
\centering
\includegraphics[height=0.50\textwidth]{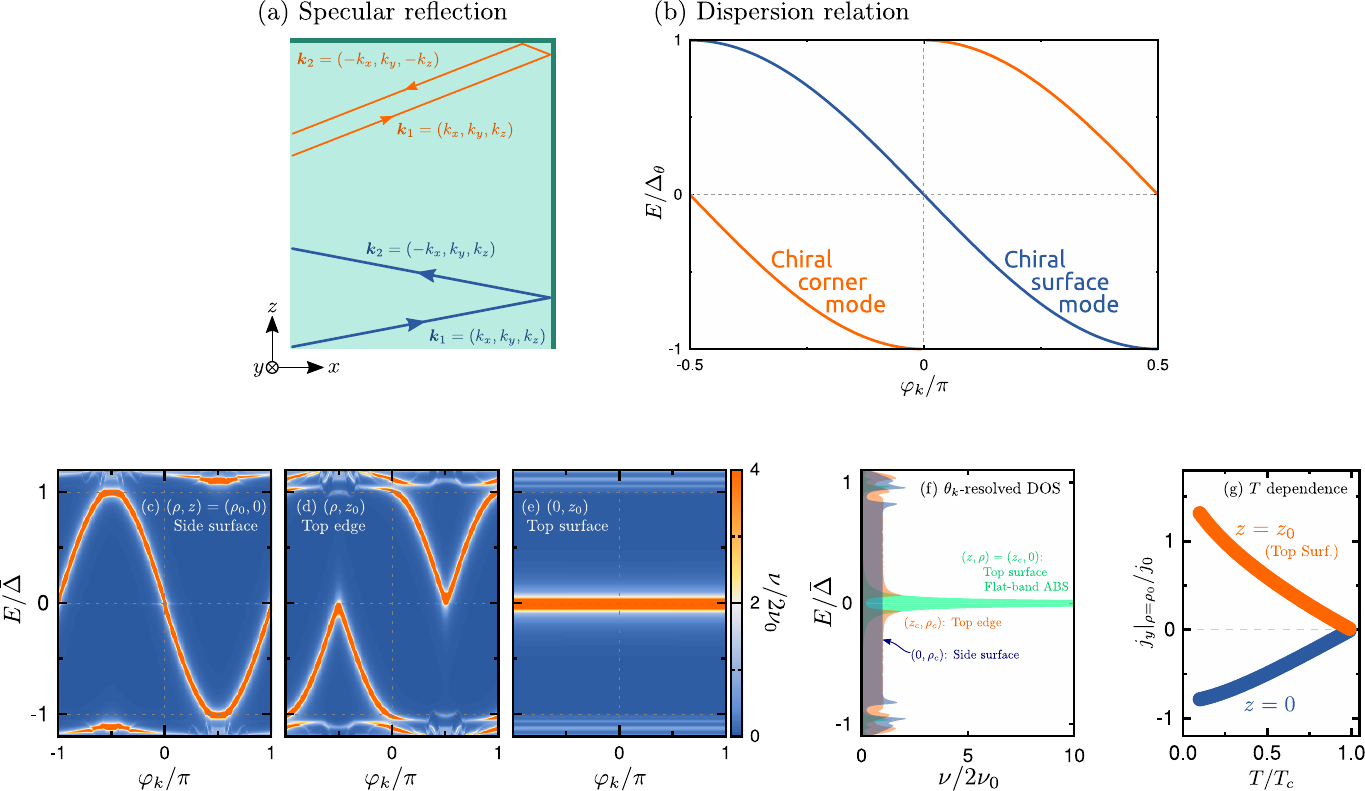}
\caption{\textbf{(a)} Schematics of edge and lateral-surface specular reflections. 
\textbf{(b)} Dispersion relations for the chiral surface state at the
lateral surface and the bound state at the top edge of the disk. 
\textbf{(c-e)} Angle-resolved density of states with the non-self-consistent
pair potential. The ARDOSs are calculated at 
(c) $(\rho, z) = (\rho_0, 0)$, 
(d) $            (\rho_0, z_0)$, and 
(e) $            (     0, z_0)$ with 
the fixed angle $\theta_k = \pi/4$. The dispersion relations are very
different from those with the self-consistent pair potential
(see main text). 
\textbf{(f)} $\theta_k$-resolved DOS with $\theta_k = \pi/4$. Under
the homogeneous pair potential, the
zero-energy peak does not appear at the top edge. 
\textbf{(g)} Temperature dependence of the chiral surface current 
($z = 0$) and 
the chiral counter current ($z = z_0$) with the homogeneous pair potential. }
\label{fig:disp}
\end{figure*}

The QGF at the surface can be obtained from the coherence
functions. The non-self-consistent coherence functions are given by 
\begin{align}
  & \bar{\gamma}_{\bs{k}}
	= \frac{i \Delta_{\bs{k}} }{ \ve + \Omega_{\bs{k}} }, 
	\hspace{12mm}
	  \bar{\ut{\gamma}}_{\bs{k}}
		= \frac{ \ve - \Omega_{\bs{k}} }{i \Delta_{\bs{k}} }, 
\end{align}
where $\Omega_{\bs{k}} = \sqrt{\ve^2 - |\Delta_{\bs{k}}|^2}$. The
specular reflection at the surface changes the quasiparticle momentum
from $\bs{k}_1 = (k_x, k_y, k_z)$ to $\bs{k}_2 = (-k_x, k_y, k_z)$,
where we have assumed the SC occupies $x \leq 0$. The schematic of the
quasiparticle reflections are shown in Fig.~\ref{fig:disp}(a). The QGF
at the surface is given by 
\begin{align}
	g^R (\bs{r}, \bs{k}; E)|_{x=0}
	= \frac{1 + \gamma_{\bs{k}_1} \ut{\Gamma}_{\bs{k}_1}}
	       {1 - \gamma_{\bs{k}_1} \ut{\Gamma}_{\bs{k}_1}}
	= \frac
	{1 + \gamma_{\bs{k}_1} \ut{\gamma}_{\bs{k}_2}}
	{1 - \gamma_{\bs{k}_1} \ut{\gamma}_{\bs{k}_2}}, 
\end{align}
where $\Gamma$ and $\ut{\Gamma}$ are the outgoing coherence functions
and the bar accents are omitted. 

We first focus on the bound states appearing at the lateral surface. The 
pair potentials for the ($p+ip'$)- and ($d+id'$)-wave SCs are given by 
$\Delta_{\bs{k}} = \Delta_0 \sin \theta_k e^{i \vphi_k}$ and 
$\Delta_{\bs{k}} = \Delta_0 \sin (2\theta_k) e^{i \vphi_k}$. Therefore, 
the pair potential before and after the reflection can be written as 
\begin{align}
  \Delta_{\bs{k}_1} = \Delta_\theta e^{i \vphi_k}, 
	\hspace{8mm}
  \Delta_{\bs{k}_2} 
	= \Delta_\theta e^{ i (\pi-\vphi_k)}
	=-\Delta_\theta e^{-i \vphi_k}, 
  \label{}
\end{align}
where $\Delta_\theta = \Delta_0 \sin \theta$ for the ($p+ip'$)-wave SC and 
$\Delta_\theta = \Delta_0 \sin (2\theta)$ for the ($d+id'$)-wave SC.
Hereafter, we make the subscript $k$ explicit only when necessary. 
The normal Green's function at the surface becomes 
\begin{align}
	g^R
	& = \frac
	{\Omega_0 \cos \vphi - i \ve      \sin \vphi}
	{\ve      \cos \vphi - i \Omega_0 \sin \vphi}. 
\end{align}
where we have used 
\begin{align}
  \gamma_{\bs{k}_1} \ut{\gamma}_{\bs{k}_2}
	=
	-
	\frac
	{(\ve-\Omega_0) e^{+i \vphi}}
	{(\ve+\Omega_0) e^{-i \vphi}}
  \label{eq:GF01}
\end{align}
and $\Omega_\theta = \sqrt{\ve^2-|\Delta_\theta|^2}$. 

Assuming $|E|<\Delta_0$, the QGF can be further reduced as 
\begin{align}
	g 
	& = i \frac
	{(\Omega'C - ES) -i\delta (EC/\Omega' + S)}
	{(EC+\Omega'S) + i \delta (C - ES/\Omega')}. 
	%
\end{align}
where we have used $\ve = E+ i\delta$, 
we have 
$
  \Omega_\theta
	\approx i  
	\left( \Omega' - i \delta {E}/{\Omega'}\right)
$, 
$\Omega' = \pm \sqrt{ \Delta_\theta^2 - E^2 }$, 
and have introduced the notations: $S = \sin \vphi$ and $C = \cos
\vphi$. 
The DOS is calculated from the real part of the QGF, 
\begin{align}
	N(E)
	%
	%
	& =  \frac
	{\delta \Delta^2 / \Omega'}
	{(EC+\Omega'S)^2 + \delta^2 (C - ES/\Omega')^2}. 
\end{align}
Since the DOS should be positive, the sign of $\Omega'$ is positive; 
$\Omega' = + \sqrt{\Delta_0 - E^2}$. 
We finally reach the DOS for the chiral surface states, 
\begin{align}
	N(E)
	& =  
	\frac{\delta \Delta^2}{\Omega}
	\frac
	{1}
	{(EC+\Omega S)^2 + \delta^2 (C - ES/\Omega)^2}. 
\end{align}
The position of the peak in the DOS can be obtained by analysing the
denominator. When the denominator becomes zero (i.e.,
$EC+\Omega'S=0$), the DOS has a peak at $E=-\Delta_0 \sin \vphi$.
This solution is consistent with the results [M.~Matsumoto and and
M.~Sigrist, {J. Phys. Soc. Jpn. \textbf{68} 3 (1999)}. 
A.~Furusaki, M.~Matsumotoand and M.~Sigrist,
\href{https://link.aps.org/doi/10.1103/PhysRevB.64.054514}{Phys. Rev.
B \textbf{64}, 054514 (2001)}]. The obtained dispersion relation is shown in
Fig.~\ref{fig:disp}(b). 

When a quasiparticle is reflected at an edge, the momentum changes
from  $\bs{k}_1 = (k_x, k_y, k_z)$ to $\bs{k}_2 = (-k_x, k_y, -k_z)$
[see Fig.~\ref{fig:disp}(a)]. In this case, we may have an unusual
edge state. We here consider the edge state of the ($d+id'$)-wave SC
whose pair potential is given by 
\begin{align}
\Delta_{\bs{k}}
= 2 \Delta_0 (k_x + ik_y) k_z = \Delta_0 \sin (2 \theta) e^{i \vphi}, 
\label{}
\end{align}
where the factor 2 is introduced so that
$\mathrm{max}[\Delta_{\bs{k}}]=\Delta_0$. While the reflection, the
pair potential changes as follows, 
\begin{align}
  \Delta_{\bs{k}_1} = \Delta_{\theta} e^{i \vphi}, 
	\hspace{8mm}
  \Delta_{\bs{k}_2} 
	=-\Delta_{\theta} e^{i (\pi-\vphi)}
	= \Delta_{\theta} e^{-i \vphi}. 
  \label{}
\end{align}
Accordingly, 
\begin{align}
  \gamma_{\bs{k}_1} \ut{\gamma}_{\bs{k}_2}
	=
	+
	\frac
	{(\ve-\Omega_{\theta}) e^{+i \vphi}}
	{(\ve+\Omega_{\theta}) e^{-i \vphi}}, 
  \label{eq:GF02}
\end{align}
where $\Omega_{\theta} = \sqrt{\ve^2 - \Delta_{\theta}^2}$. 
Equation \eqref{eq:GF02} has an extra $\pi$ phase compared with
Eq.~\eqref{eq:GF01}. This extra phase comes from the edge reflection. 
Hereafter, we make the subscript $\theta$ explicit only when
necessary.  The QGF becomes, 
\begin{align}
	g 
	& = \frac
	{\ve     C - i \Omega' S}
	{\Omega' C - i \ve     S}
	\\
	& = -i \frac
	{(EC + \Omega' S) +i \delta (C-ES/\Omega')}
	{(\Omega' C - ES) -i \delta (EC/\Omega' +S )}. 
\end{align}
The real part of the QGF is 
\begin{align}
	N(E) 
	& = \frac{\delta \Delta_\theta^2}{\Omega'} \frac
	{1}
	{(\Omega' C - ES)^2 + \delta^2 (EC/\Omega' +S )^2}, 
\end{align}
leading to $\Omega' = \Omega = \sqrt{\Delta_\theta^2-E^2}$.  The DOS
for the edge states is 
\begin{align}
	N(E) 
	& = \frac{\delta \Delta_\theta^2}{\Omega} \frac
	{1}
	{(\Omega C - ES)^2 + \delta^2 (EC/\Omega +S )^2}. 
\end{align}
The DOS diverges at $E = \cos \vphi \sin \vphi_k / |\sin \vphi_k|$. 
The dispersion relations for the chiral surface states and the edge
states are shown in Fig.~\ref{fig:disp}(b). 

The ARDOS for a homogeneous pair potential is shown in
Fig.~\ref{fig:disp}(c-e), where the results are calculated at 
(c) $(\rho,z)=(\rho_c,  0)$ (i.e., lateral surface), 
(d)          $(\rho_c,z_c)$ (i.e., top edge), and 
(e)          $(   0,z_c)$ (i.e., top surface) and the other 
parameters are the same as in Fig.~\ref{fig:ARDOS_main}. The chiral
surface states and the flat-band ABSs have qualitatively the same
dispersion relation as those obtained with the self-consistent pair
potential [Figs.~\ref{fig:ARDOS_main}(a,e) and \ref{fig:disp}(c,e)].
The chiral edge states, on the other hand, show different dispersion
relation compared with the results in the self-consistent simulations
[Figs.~\ref{fig:ARDOS_main}(c) and \ref{fig:disp}(d)]. In the
non-self-consistent simulation, the edge bound states appear only in
the regions with $E \sin \phi_k > 0$ where the bound states carry the
electric current in $+y$ direction.  In addition, the height of the
ARDOS is also influenced by the self-consistency. The
$\theta_k$-resolved DOS with a homogeneous pair potential is shown in
Fig.~\ref{fig:disp}(f). Figure~\ref{fig:disp}(f) shows that no
zero-energy peak appears at the top edge. The zero-energy peak of the
chiral edge states play an important role in the temperature
dependence of the CCC. Therefore, with the non-self-consistent pair
potentials, the CCC does not show the low-temperature enhancement
[Fig.~\ref{fig:disp}(g)]. We can conclude the self-consistency is
important to analyze the CCC. 

\begin{figure*}[tb]
\centering
\includegraphics[height=0.25\textwidth]{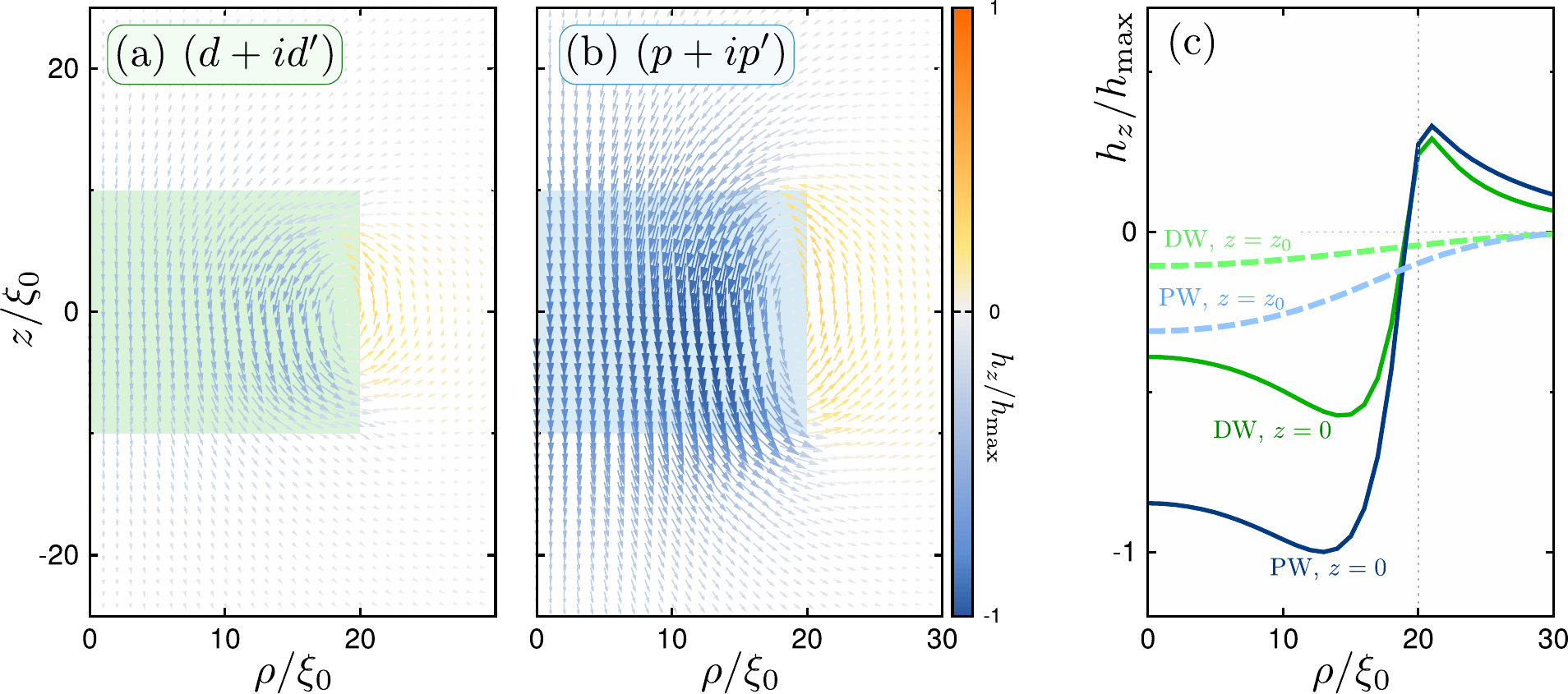}
\caption{
  Vector plot of magnetic field for $z_0 = 10\xi_0$ for 
  \textbf{(a)} $(d+id')$-wave and \textbf{(b)} $(p+ip')$-wave SCs. 
  The spontaneous magnetic field in the $(d+id')$-wave SC is smaller
  than that in the $(p+ip')$-wave SC because of the counter chiral 
  current at the top/bottom edges. \textbf{(c)} $z$-component of the
  magnetic field. The results at $z=0$ (middle of the disk) and $z=z_0$
  (top surface) are shown. 
  The contribution from the edge current is smaller than that the
  results for $z_0 = 5\xi_0$. The temperature is fixed at $T=0.2T_{c0}$. }
\label{fig:field_thick}
\end{figure*}

\section{Spontaneous magnetic field of a thick disk}

The vector plots of the magnetic field $\bs{H}(\bs{r})$ for 
the ($d+id'$)- and ($p+ip'$)-wave SCs are shown in
Fig.~\ref{fig:field_thick}(a) and \ref{fig:field_thick}(b), respectively. 
The $z$-components of the magnetic fields $H_z(\bs{r})$ are plotted in
\ref{fig:field_thick}(c), where $H_z(\bs{r})$ are normalized to 
$H_{\mathrm{max}} \equiv \mathrm{max}[H_z(\bs{r})]$ in the ($p+ip'$)-wave SC.
In Fig.~\ref{fig:field_thick}(c), we obtained the results at 
$z=z_0$ (top surface of the disk) and $z=0$ (middle of the disk). 
With increasing the thickness, the area where the chiral surface
current flows increases. However, $H_z$ at the top surface of the 
($d+id'$)-wave disk remains much smaller than that in the ($p+ip'$)-wave
disk. Although the area for the chiral counter current is smaller
compared with that for the chiral surface current, the contribution
from the chiral counter current is comparable with that from the
chiral current because the chiral counter current is much closer to
the top surface [See Eq.~\eqref{eq:BS}]. 

\section{Chiral current density of a thin disk}

\begin{figure*}[t]
\centering
\includegraphics[height=0.25\textwidth]{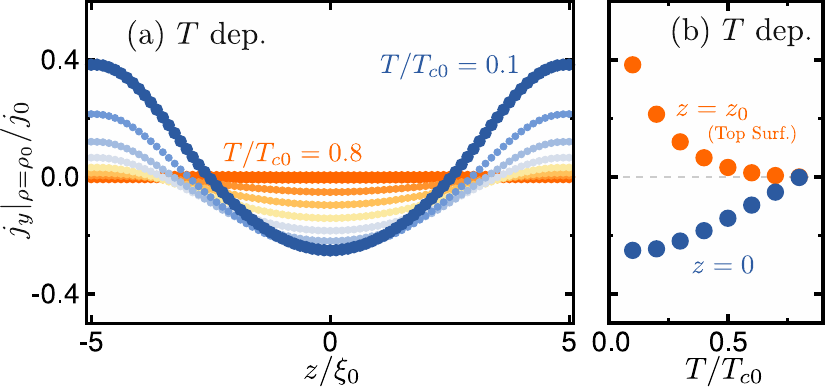}
\caption{
\textbf{(a)} Temperature dependence of the chiral surface
current in ($d+id$)-wave SC with $z_0=5\xi_0$. \textbf{(b)} Amplitude
of the current density at $z=0$ and $z=z_c$.  The parameters are 
$\rho_0=20\xi_0$ and $T=0.2T_{c0}$. }
\label{fig:cur_thin}
\end{figure*}

The current density at the lateral surface ($\rho=\rho_0$) is shown in 
Fig.~\ref{fig:cur_thin}(a), where the temperature $T$ varies from
$T=0.2T_{c0}$ to $0.8T_{c0}$ by $0.1T_{c0}$. The results at $z=z_0$
(top edge) and $z=0$ (middle of the disk) are shown in
Fig.~\ref{fig:cur_thin}(b). The temperature dependences are
qualitatively the same as those for $z_0=10\xi_0$: the chiral counter 
current increases at low temperatures. 

\end{document}